\documentclass[11pt,nofootinbib,showpacs]{revtex4}
\usepackage{amsfonts,amssymb,amsmath,graphicx}
\def\[{\left[}
\def\]{\right]}
\def\({\left(}
\def\){\right)}

\begin{document}

\title{Power-law tails in triple system decay statistics}

\author{A. V. Bogomolov}
\affiliation{Skobeltsyn Institute of Nuclear Physics, Moscow State
University, Moscow, 119991 Russia}

\author{S. A. Pavluchenko}
\affiliation{Special Astrophysical Observatory, Russian Academy of
Sciences, Nizhnij Arkhyz, 369167 Russia}

\author{A. V. Toporensky}
\affiliation{Sternberg Astronomical Institute, Moscow State
University,Moscow, 119992 Russia}

\begin{abstract}
We have investigated the decay statistics of triple systems with
different masses in Newtonian dynamics. We  demonstrate that in a
broad interval of mass ratios this statistics has good
approximation by power-law tails.   The power indices do not show
any significant dependence on mass ratios.
\end{abstract}

\maketitle \twocolumngrid

\section{Introduction}

Power-law tails in various branches of natural science have been
intensively investigated during last several decades. These
studies deal with experimental results on existing of power-law
tails as well as attempts of theoretical explanation of this
phenomenon. Some experimental material have been collected in a
popular review \citep{FCP}, other examples important for
astrophysics see, for example, in \cite{a1, a2}. As an example of
theoretical description it is possible to note a progress in
modern statistical physics which predicts power-law tails by using
non-classical definitions of entropy  (this approach starts from
works of \cite{R} and \cite{T}, for a modern review see, for
example, \cite{Rudoi}). For other approaches to possible source of
this phenomenon see, for example, \cite{Montroll}.

In hamiltonian dynamics power-law tails appear in two different
situations: for systems with escape we can study the number of
trajectories $N$ which survive up to time $t$ (or, alternatively,
the number of trajectory $dN$ experienced the escape in the time
interval from $t$ to $t+dt$ we will study this differential form
of the time distribution of escapes in the present paper), for
systems with a compact phase space it is possible to construct a
function which describes Poincare recurrence time depending on the
initial position in the phase space. Many empirical results show
that the appearance of power-law tails in such systems is
connected with islands of a regular motion in the phase space and
``stickness'' of trajectories to the boundaries of these islands
\citep{Zaslavsky}, though strict mathematical results about this
kind of systems are still rather poor.

Three-body problem is an example of combined dynamics -- it is
known that stable periodic orbits surrounded by islands of regular
dynamics exist in a chaotic "sea" representing the bulk of phase
space. For equal-mass problem (when all three masses have the same
value) there exists stable configuration called a ``figure-eight''
orbit \citep{fe}, as well as several other orbits \citep{Orlov}.
It is also known that the "figure-eight" orbit becomes unstable
when relative differenses in masses reach approximately $10^{-2}$
\citep{Simo}, so this very interesting orbit is hardly important
for any real astrophysical problem. The Schubart orbit \citep{s}
remaines stable in a much broader interval of masses
\citep{Shuborb}
 On the other hand, famous triangle Lagrange
points become stable in the situation of suffitiently different masses, and we can imagine that
set of stable orbits have rather nontrivial dependence on mass differences.

The phenomenon of trajectory ``stickness'' to boundary of a
regular region for the three-body problem has already been
remarked by \cite{st} and \cite{Orlov}. Keeping this in mind we
can expect existence of power-law tails in  $N(t)$ distribution.
For other theoretical arguments supporting this suggestion see
\cite{Agekian}. Earlier attempts to describe this function have
been done by \cite{Mikkola} where  exponential ansatz for $N(t)$
have been used. However, Orlov et al. indicated that this
modelling works only for a limited time interval, and for large
enough time $t$ power-law tails become clearly detectable
\cite{Orlnew}. Both these works deal with the equal-mass problem.
This construction may be too symmetric for making general
conclusions (keeping in mind loss of stability for "figure-eight"
solution" with a small mass inequality), so it is reasonable to
provide similar analysis in a general situation with non-equal
masses. In the present paper we have chosen several different mass
ratios for investigation, starting from near-equal mass system and
concluding for the system with masses $(0.1, 1, 10)$ mass units,
which means that Schubart orbit is already unstable, and Lagrange
triangle points are still unstable for this mass ratio.

\section{System and initial data}

\begin{figure}
\includegraphics[width=8cm]{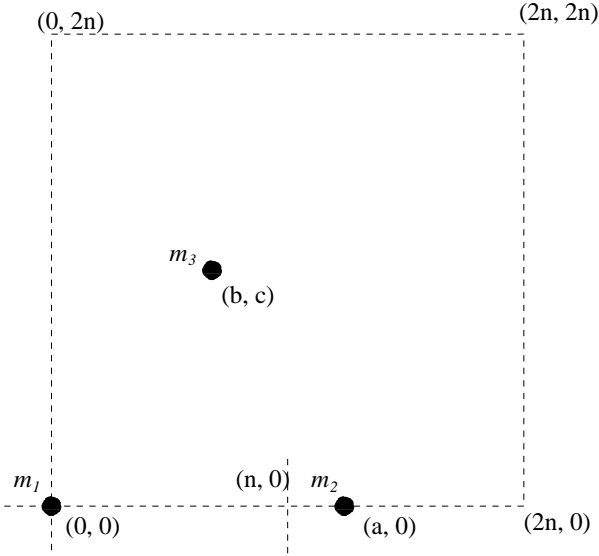}
 \caption{The layout of the initial conditions (coordinates). The $m_1$ body is always in the (0, 0)
 point, $x$-coordinate of $m_2$ is varying within certain range. As for $m_3$, its both coordinates
 are varying --- see text for details.}
\end{figure}

In the present paper we cover wide range of mass ratios from
equal-mass problem to near hierarchical system. We use Aarseth
code \citep{Aarseth} and start from zero-velocity initial
conditions. Time variable is chosen so that $G=1$.

The layout of the initial conditions we used in our study is given in
Fig. 1. The first body  with mass $m_1$ is initially located in the
(0, 0) point, the second body $m_2$ initially lies on the ($a$, 0)
segment with $a\in (n+\delta, 2n+\delta)$, where $n$ is a space scale
and $\delta$ is some small separation introduced to avoid
the situation when two bodies have totally coinciding initial coordinates. Finally, the third body
$m_3$ initially lies within a square $(b, c)$ with $b\in (0, 2n)$ and
$c\in (0+\delta, 2n+\delta)$.
This choice cover rather wide area in the initial condition space for our results to
be representent, on the other hand, different permutations of non-equal masses
results in different initial condition sets allowing us to study possible dependence
of our results upon initial conditions chosen.

Using the described above layout we evenly distribute about 4
millions initial configurations and calculate their evolution
until the decay (for exact meaning of the decay in Aarseth code
see, for example, \cite{Orlnew}).

\begin{figure}
\includegraphics[angle=-90, width=9cm]{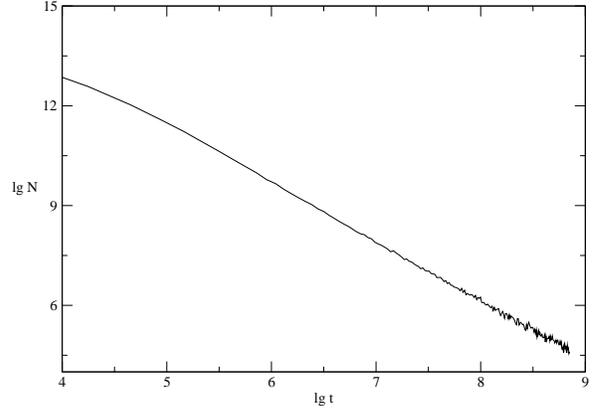}
 \caption{The typical plot of $N(t)$ in the logarithmic scale.}
\end{figure}

\section{Decay spectra}

A typical plot of the differential distribution $N(t)$ in the
logarithmic scale is shown in Fig. 2. Power-law tail following a
steeper function after $t$ reaches the value of about one thousand
is clearly seen. The distribution of escaping time for small $t$
is clearly different from power-law, and to obtain correct values
of power index in the modelling
$$
N(t)= {\rm const} \times t^{-\gamma}
$$
we should exclude this range of $t$ from
the analysis. As an illustration we plot in Fig. 3 power indices calculated in the time range from
 $t_{in}$ to $T_f$ where the maximal time of integration $T_f=7000$. We
can see that the power index typically does not change if $t_{in}$
is bigger than $2000$. In what follows we will ignore first $2000$
time units in calculation of power index $\gamma$. In the table
below we present minimal and maximal values of $\gamma$ for
several mass combinations calculated for different sets of initial
conditions. Error bars, also presented in the table indicate that
in some situations the difference in power indices calculated for
the same system using different sets of initial conditions can be
real (though not so big). We  can see also that despite mass ratio
of the systems under investigation cover a broad interval from
near-equal to near-hierarchical systems, the value of $\gamma$
does not change significantly with the slightly developed tendency
of growing $\gamma$ with the increasing hierarchicity of the
triple system. We should, however, note that our values of
$\gamma$ for equal mass system is bigger than the value found in
the paper \cite{Orlnew}. Can different sets of initial conditions
be the only cause of this difference remaines unclear and requires
further investigations.

\begin{figure*}
\includegraphics[angle=-90, width=14cm]{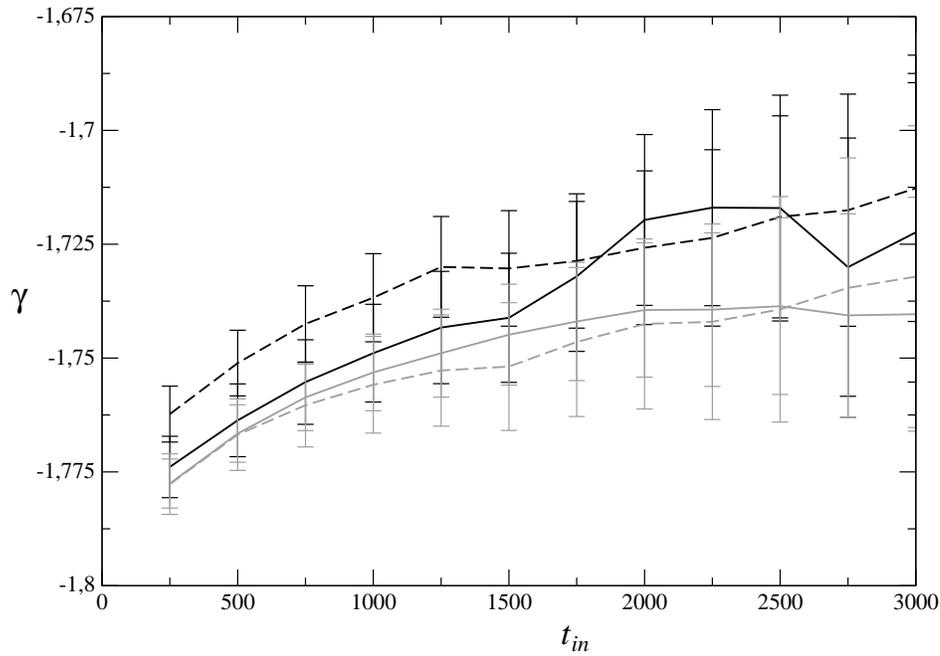}
 \caption{An examples of dependence of the slope $\gamma$ on the cutoff time
 $t_{in}$. Black lines corresponds to (1, 2, 3)$_{n=3}$ mass combination, grey -- to
 (1, 3, 5)$_{n=5}$. Different line styles correspond to different distribution of masses between
 $m_1$, $m_2$ and $m_3$.}
\end{figure*}

\begin{table}
 \centering
 \begin{minipage}{70mm}
  \caption{Maximal and minimum values for the slope $\gamma$, calculated at $t_{in}=3000$ for different mass
  combinations. First column gives absolute masses, second -- masses, normalized
  to $m_1 + m_2 + m_3 = 1$, third -- the slope $\gamma$ with an error, minimal (upper row) and maximal (lower row) values.}
  \begin{tabular}{@{}cccc@{}}
  \hline
  \multicolumn{2}{c}{Masses}  &  \multicolumn{2}{c}{$\gamma \pm \delta\gamma$}   \\
 \hline
(0.1, 1, 10) & (0.009, 0.09, 0.901) &   1.762 & 0.011  \\
             & &   1.863 & 0.016  \\
\hline
(1, 3, 10)   & (0.071, 0.214, 0.715) &  1.845 & 0.013  \\
             & &   1.862 & 0.012  \\
\hline
(1, 3, 5)    & (0.111, 0.333, 0.556) &  -1.828 & 0.013  \\
             & &   1.848 & 0.014   \\
\hline
(1, 2, 3)    & (0.167, 0.333, 0.5) &  1.857 & 0.014  \\
             & &   1.905 & 0.015  \\
\hline
(0.9, 1, 1.1)    & (0.3, 1/3, 11/30) &  2.149 & 0.009   \\
             & &   2.181 & 0.010   \\
\hline
(1, 1, 1)    & (1/3, 1/3, 1/3) &  2.125 & 0.014   \\
             & &   2.178 & 0.012   \\
\hline
\end{tabular}
\end{minipage}
\end{table}

\section{Conclusions}

We studied the escaping rate statistics in a general three-body
problem with non-equal masses. We argue that for large enough time
$t$ this statistics can be modelled by power-law functions with a
good accuracy. For all cases studied (with normalized mass ratios
from to) the power index is located within a rather narrow
interval (from $1.76$ to $2.19$).

\section*{Acknowledgments}

Authors are grateful to Artur Chernin, Viktor Orlov and Sergey
Prants for helpful discussions. As well as we'd like to thank
Sergey Karpov (SAO RAS) for useful discussions and various
technical help.

\onecolumngrid

\end{document}